\documentclass[runningheads]{llncs}
\usepackage{graphicx}
\usepackage{tikz,xcolor,hyperref}

\definecolor{lime}{HTML}{A6CE39}
\DeclareRobustCommand{\orcidicon}{
\begin{tikzpicture}
\draw[lime, fill=lime] (0,0)
circle[radius=0.16]
node[white]{{\fontfamily{qag}\selectfont \tiny \.{I}D}}; 
\end{tikzpicture}
\hspace{-2mm}
}
\foreach \x in {A, ..., Z}{%
\expandafter\xdef\csname orcid\x\endcsname{\noexpand\href{https://orcid.org/\csname orcidauthor\x\endcsname}{\noexpand\orcidicon}}
}

\usepackage{algorithm}
\usepackage{algorithmic}
\usepackage{float}
\usepackage{multirow}
\usepackage{booktabs} 
\usepackage{hhline}

\usepackage[utf8]{inputenc}
\usepackage{pifont}
\usepackage{newunicodechar}
\usepackage{rotating}

\usepackage{wrapfig}

\usepackage{adjustbox}
\newunicodechar{✓}{\ding{51}}
\newunicodechar{✗}{\ding{55}}

\usepackage{multirow}
\usepackage{marvosym}

\begin{document}
\title{Adaptive White-Box Watermarking with Self-Mutual Check Parameters in Deep Neural Networks}
%
%


%
%

\author{Zhenzhe Gao\inst{1}\hspace{-1.5mm}\orcidA{} \and
Zhaoxia Yin\inst{1}\orcidB{} \and
Hongjian Zhan\inst{1,2}\hspace{-1.5mm}\orcidC{} \and
Heng Yin \inst{3}\hspace{-1.5mm}\orcidD{} \and
Yue Lu \inst{1}\textsuperscript{({\normalsize{\Letter}})}\hspace{-1.5mm}\orcidE{}}



\institute{Shanghai Key Laboratory of Multidimensional Information Processing, East China Normal University, Shanghai 200241, China 
\\ 
\email{ylu@cs.ecnu.edu.cn}\\ \and Chongqing Institute of East China Normal University. Chongqing. 401120. China  \and
Anhui Provincial Key Laboratory of Multimodal Cognitive Computation, Anhui University, Hefei 230000, China}

\maketitle              

\begin{abstract}
Artificial Intelligence (AI) has found wide application, but also poses risks due to unintentional or malicious tampering during deployment. Regular checks are therefore necessary to detect and prevent such risks. Fragile watermarking is a technique used to identify tampering in AI models. However, previous methods have faced challenges including risks of omission, additional information transmission, and inability to locate tampering precisely. In this paper, we propose a method for detecting tampered parameters and bits, which can be used to detect, locate, and restore parameters that have been tampered with. We also propose an adaptive embedding method that maximizes information capacity while maintaining model accuracy. Our approach was tested on multiple neural networks subjected to attacks that modified weight parameters, and our results demonstrate that our method achieved great recovery performance when the modification rate was below 20\%. Furthermore, for models where watermarking significantly affected accuracy, we utilized an adaptive bit technique to recover more than 15\% of the accuracy loss of the model.

\end{abstract}
\section{Introduction}
\begin{figure}[htp]
	\centering
		\includegraphics[scale=.3]{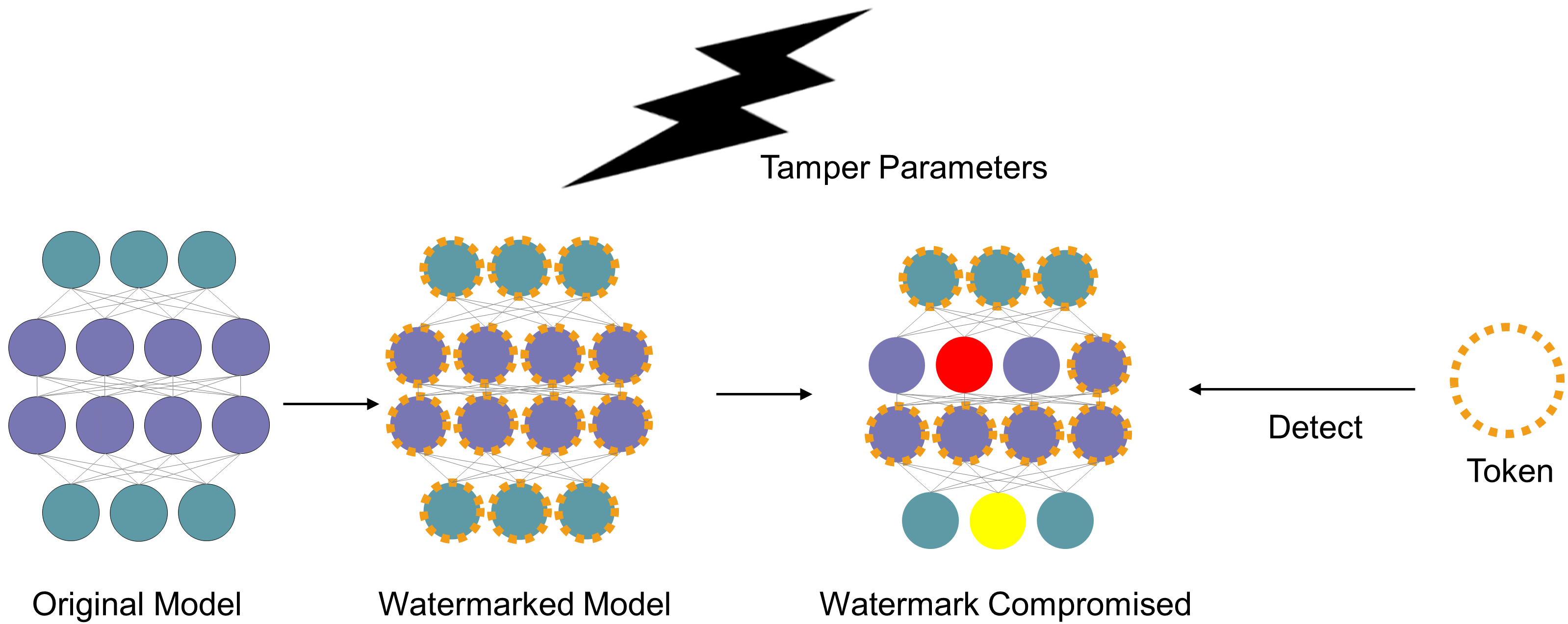}
	\caption{Model parameters can be tampered with, and fragile watermarks can establish a fragile barrier that allows users or model owners to check the status of model parameters through tokens.
}
	\label{figure1}
\end{figure}

Deep neural networks (DNNs) are often deployed in various fields, such as image classification \cite{imgrec1}and natural language processing \cite{nlp1}. Due to the varying sizes of neural network models, we deploy artificial intelligence models on cloud \cite{ribeiro2015mlaas} or embedded devices \cite{embedding}. Regardless of the deployment method, it is challenging for users to ensure that the model is fully deployed as intended by the owner. The model can be subjected to quantization or pruning to reduce server load, and it may also be vulnerable to attacks that modify the model parameters, such as backdoor attacks or poisoning attacks \cite{attack2019survey,chen2018detecting_backdoorandpoison}. By embedding watermark information into the parameters, it serves as a fragile barrier for the model parameters, allowing us to determine whether the parameters have been tampered with by examining the parameters themselves, as shown in Figure \ref{figure1}. 
\vspace{-.5em}

Although standard methods for data integrity checks, such as SHA-256 \cite{sha-256} and CRC \cite{CRC}, exist, adjustments need to be made to the calculation method of the password in different in model frameworks. Additionally, because of the characteristics of neural network models and hash, it is powerless to locate and recover tampering.

Model watermarking technology \cite{li2021survey} is a technique that combines the characteristics of neural network models, used for protecting model intellectual property and model integrity. Intellectual property was the first application of watermarking when neural network watermarking are proposed by \cite{first}. The watermark for integrity protection is often referred to as a fragile watermark and is currently roughly divided into two directions: black-box fragile watermark and white-box fragile watermark.
Fragile watermarking refers to the ability of the watermark to reflect any modifications made to the model, thereby determining its integrity status. Black-box fragile watermarking assumes that the model can only be queried through its input and output interfaces, and by testing specific inputs (triggers, also known as sensitive samples), it is possible to determine if the model has been tampered with.
There have been many previous works in this field, including the most representative work by He et al. \cite{sensitive} from Princeton University, who used the Taylor series expansion of the neural network to describe the formula for attacking the neural network, and found the most sensitive sample that could best reflect the small changes in the neural network as the sensitive sample. Kuttichira et al. \cite{pys} searched for specific triggers by building an optimizer suitable for Bayesian algorithms, and achieved detection against any attacks in experiments, but the detection efficiency was not high. O. Aramoon et al. \cite{aid} believed that triggers that fall on the classification boundary are the required triggers for classification tasks, but for other tasks, the model's decision boundary is not as easily constructed based on output probabilities as in classification tasks. Yin et al. \cite{yin2022neural} used a generative adversarial nets\cite{gan} to learn the model's boundaries and generate sensitive samples autonomously.

The aforementioned black-box model watermarking techniques are limited in their detection capabilities due to their pre-defined API-based approach. Due to the opacity of neural networks, it is challenging to be certain that black-box methods can detect all potential attacks with 100\% accuracy. Furthermore, it is difficult to achieve localization and recovery. Therefore, white-box watermarking is necessary as a more rigorous approach to be applied in neural networks. White-box fragile watermarks allow for viewing of the model's internal parameters. However, this does not mean that one can easily obtain the true original model for comparison, as on the cloud, it is difficult to distinguish between the original model and its tampered copy. And for offline devices, it is even more difficult to conduct online comparison. Previous work on white-box fragile watermarks includes Li et al. \cite{radar} who studied the attack patterns of the PBFA algorithm for specific neural networks, and placed carefully designed model parameter check bits on a separate memory to detect model integrity at runtime. Additionally, they leveraged the technique of setting erroneous block parameters to zero in order to restore model performance. Botta et al. \cite{neunac} achieved block-level positioning by using KL transforms and genetic algorithms to set the least significant bits (LSBs) of the parameters as watermark bits, but this still resulted in model performance degradation and it causes detection omissions. Similarly, Zhao et al. \cite{zhao2022dnn} from University of Shanghai for Science and Technology introduced the self-embedding technique used in image fragile watermarking to DNNs, setting the 12 LSBs of the neural network parameters as watermark bits, achieving 100\% detection of neural network tampering, block-level positioning, and partial recovery of neural network performance, similar to recovery in the image domain.

In our approach, we scrambled the parameters using a specific permutation and placed the important information of the previous parameter in the position of the unimportant information of the subsequent parameter. Meanwhile, we used mod operation on each parameter itself to achieve precise detection, accurate localization, and precise recovery at the parameter level. Parameters of neural networks differ from those of images in several ways. For instance, high-frequency features in images are sensitive to human perception, and therefore need to be protected when embedding watermarks. However, the importance of parameters to the results in neural networks is related to the gradients, magnitude, and position of tensors. As neural networks become deeper, even small changes to individual parameters can have a significant impact on the final output, rendering the traditional image watermarking inapplicable. Similarly, the Peak Signal-to-Noise Ratio (PSNR) \cite{psnr} commonly used in the image field to indicate the similarity between images is not a incompatible indicator of model variation in neural networks. Moreover, for white-box watermarks, we often need to replace the least significant bits (LSBs), but the watermarking that have little impact on small models can lead to a significant performance decrease when placed on deep neural models. Therefore, we have developed an adaptive bit adjustment technique that achieves a watermarking embedding capacity far greater than that of previous works. Our contributions can be summarized as follows:

\begin{itemize} \item We propose a method for generating adaptive bits based on gradient descend, which provides a way to recover the performance of the model up to 15\% when adjusting the LSBs of the model.\item Our watermarking algorithm combines the relationship of parameters and the relationship among the parameter's own bits, achieving 100\% detection of model modification, parameter-level positioning of tampered regions, and recovery of model performance for modifications below 20\%.  \item We conduct a comparative analysis with previous integrity verification methods, demonstrating that our approach is the first to achieve precise parameter-level localization while preserving the original performance of the model.  
\end{itemize}

\section{Adaptive Watermarking}

\subsection{Problem Formulation}
For methods that require replacing LSBs to embed watermarks, it is desirable to minimize the change in model performance caused by LSBs for each parameter $W_{ij}$ in the neural network. In the field of image processing, PSNR is often used to describe the differences between images. However, in the field of neural networks, \cite{xue2022advparams} have shown that even a small number of parameters can have a significant impact on model performance, and PSNR may not be suitable for neural networks.

In this case, accuracy is used to describe the performance of the neural network after embedding the watermark, and designers aim to minimize the change in performance as much as possible. So the objective can be described as: $maxmize\bigl(Acc(f(X_{test},W'),Y)\bigr)$. Here, $f()$ denotes model inference and $X_{test}$ and $Y$ represent the test set and the set of labels, respectively. $Acc$ represents accuracy of the inference. $W'$ denotes the parameters with the embedded watermark. As the amount of embedded watermark content and the depth of the model increase, the method of adjusting some LSBs may also have a greater impact on the model.
\subsection{Adaptive Method}
\begin{figure}[htp]
	\centering
		\includegraphics[scale=.45]{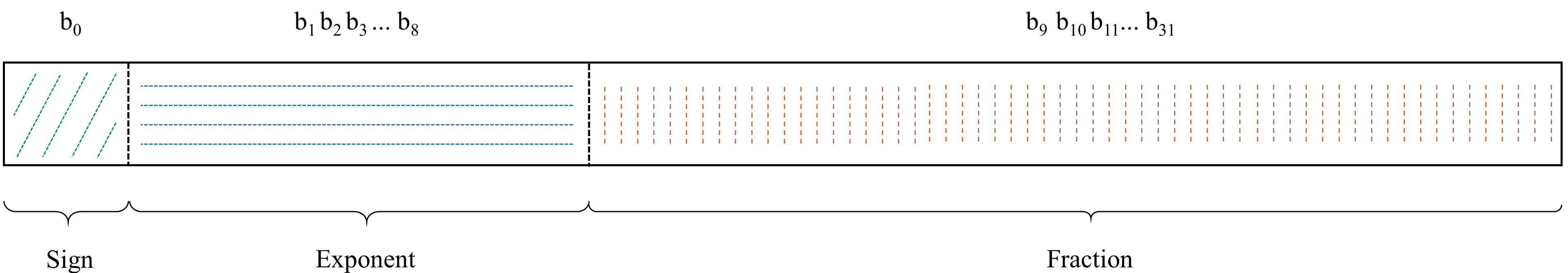}
	\caption{The sign field in IEEE 754 floating-point numbers determines the sign of the floating-point number, the exponent field determines the position of the decimal point, and the fraction field determines the significant digits.}
	\label{figure2}
\end{figure}

The existing neural network frameworks, such as Pytorch and TensorFlow, adopt default parameters that comply with the IEEE 754 protocol \cite{754} for floating-point numbers. Each floating-point number consists of 32 bits. For ease of description, we use $b_0, b_1, b_2...b_{31}$ to represent the 32 bits, where $b_0$ is the sign bit, $b_1-b_8$ are the exponent bits used to control the position of the decimal point, and the remaining bits are referred to as the fraction bits, which form the significant digits. Obviously, the value of a number is mainly influenced by the sign bit, exponent bits, and the leading fraction bits. We can gain a more intuitive understanding of Figure \ref{figure2}. In the watermark embedding method that replaces the least significant bits (LSBs), we often replace the trailing fraction bits to embed the watermark, which unavoidably causes slight changes in the original numerical values. As the neural network makes inferences layer by layer, the final results may deviate original model.


\begin{figure}[htb]
	\centering
		\includegraphics[scale=0.5]{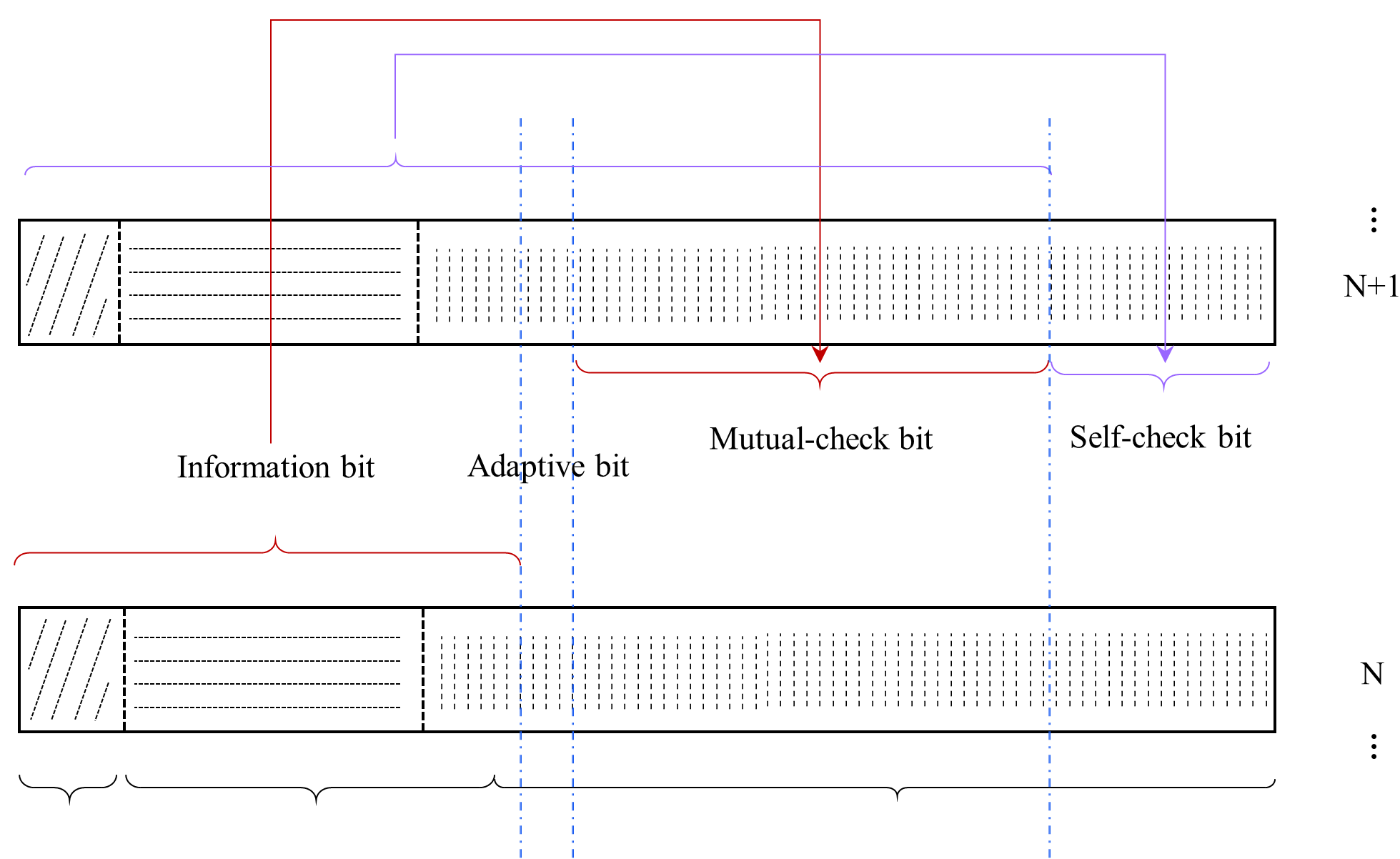}
	\caption{Construction of the parameters: Adaptive bit ($b_{11}$) is between Information bits and Mutual-self check bits.}
	\label{figure3}
\end{figure}

To address this issue, we propose an adaptive watermark embedding method. In other words, we obtain performance correction by training one bit of the parameters. Taking our watermarking method as an example, for each floating-point parameter, we need to replace the 19 least significant bits (LSBs), and therefore, we need to train the 21st bit from the end to restore performance. Intuitively, for the fraction part, the bits closer to the front have a greater impact on the value. Thus, we can correct the previous impact by influencing $b_{11}$, as shown in Figure \ref{figure3}. The generation process is described in detail in Algorithm 1. We iterate through each layer of the neural network, conduct $\alpha$ training iterations for each layer, and obtain the gradient and accuracy using the training and test sets respectively. After adjusting the watermark bits, we compare whether the accuracy has improved and save the better adjustment.

\renewcommand{\arraystretch}{1}
\setlength{\tabcolsep}{12pt}
\begin{table*}
\caption{Adjust adaptive bit in four different situations as follow.
}\label{Table:1}    
\centering
\begin{tabular}{ccc}
\hline
\textbf{Tensor} & \textbf{Grad} & $\mathbf{b_{11}}$ \\ \hline
-      & -    & 0  \\ \hline
-      & +    & 1   \\ \hline
+      & -    & 1   \\ \hline
+      & +    & 0   \\ \hline
\end{tabular}
\end{table*}
\renewcommand{\arraystretch}{1.0}

 We aim to move each parameter in the direction opposite to its gradient to achieve a decrease in the loss function, as shown in Table \ref{Table:1}. For instance, if the gradient is positive for a positive parameter, we want the tensor value to be smaller. Then we set $b_{11}$ to $0$. Similarly, if the gradient is positive for a negative parameter, we also want the tensor value to be smaller, but due to the sign, we need to make the absolute value of the tensor larger, leading to a smaller value. We set $b_{11}$ to $1$. Changes in the fraction part have adaptive capabilities due to the existence of the exponent, and we still need to control the step size of parameter changes as in normal training. 
To enhance adaptability and better control the magnitude of gradient descent, not all parameters undergo the same operation, and we define two hyper parameters, $\alpha $ and $\beta$, which represent the number of iterations for each layer and the ratio of parameters, respectively. Pseudo-code is presented below.

\begin{algorithm}
\footnotesize
\caption{Generating adaptive watermark bits}

\textbf{Input}: $\alpha$, $\beta$, neural network $f(X,W)$, where $X$ represents the input of the neural network.


\begin{algorithmic}
\FOR{Layer}
\FOR{$i$ = 1 to $\alpha$}


\STATE $Grad  \Leftarrow f(X_{train},W)$

\STATE $Acc_{o} \Leftarrow f(X_{test},W)$

\STATE Sort Tensor with Grad

\FOR{$ Pars *\beta $}
\IF{$Tensor*Grad > 0$}
\STATE  $ b_{11} = 0 $
\ELSE
\STATE $ b_{11} = 1 $
\ENDIF
\ENDFOR

\STATE $Acc_{c} \Leftarrow f(X_{test},W')$
\IF{$Acc_c > Acc_o$ } 
\STATE  $f(W)   \Leftarrow f(W')$

\ENDIF

\ENDFOR
\ENDFOR

\textbf{Output}: Adaptive neural network $f(X,W')$
\end{algorithmic}
\end{algorithm}

\section{Self-Mutual Parameter Check}
\subsection{Two Simple Assumptions}
Validation bits need to have strong sensitivity to any changes made to the model \cite{fan2019rethinking}, and this sensitivity must be related to the model itself. Although validation bits are still part of the model, watermarks hope to make the validation bits and all content of the model associated so that the model and watermark are truly and completely coupled. We can assume a scenario in which all parameters' least significant bits (LSBs) are set to 1. In this case, we can easily understand that even a tiny adjustment to the model or setting one of over millions parameter to zero (or any other value) would change the LSB of the model. However, since one is independent of other parameters of the model, we can easily implement an attack that sets all LSBs to 1, making the model's fragile watermark completely ineffective.


We can also assume another protection method: parameter backup. We select 16 bits from a 32-bit floating-point number to carry the information and another 16 bits that are exactly the same as the first 16 bits to backup and check the information bits, ensuring that any modification to the parameter causes the two 16-bit parameters to be mismatched and successfully detected, but cannot be recovered. This is because it is impossible to determine which part is incorrect. If an attacker adjust the first 16 bits and the last 16 bits to be consistent after attacking one parameter, also achieving a covert attack without being detected. Although these two examples are relatively simple, they demonstrate that the fragile watermarking needs to have a strong association with the information itself and ensure that there is a certain correlation between parameters to ensure that when modifying parameters, one must implement a tampering of all parameters to maintain the original characteristics of the watermarking while preventing attacks. Finally, it is best to add a key attribute to the fragile watermark to ensure that the watermarking information can only be obtained through the secret key.

\subsection{Constructing Self-Mutual Check Parameters}

For white-box watermarking, we aim to achieve a 100\% success rate in detecting tampering, locate the position of the tampering, and restore a certain amount of tampering. To achieve this, our design ensures the coupling of information between parameters and within individual parameters. Specifically, we designed the watermark using the following method.

The process of generating and adding the watermark is done layer by layer on a neural network. For a layer of the network, we permute its parameters with a secret key (random seed) and record the scrambling sequence for detection and restoration. After permuting, we concatenate the first and last parameters to obtain a circular sequence resembling a circle, each parameter has a parameter before and after it. To obtain the check bits, we select the first 11 bits for protection (including one sign bit, eight bits of exponent and two bits of fraction), the next bit $(b_{11})$ as the adaptive bit, and the first eight bits of the remaining 20 bits as the mutual check bit that will be determined through computation. Computation involves XOR operations between the information bits of the previous parameter, the information bits of this parameter, and the secret key (you could also increase the complexity of reversible calculations to make them more difficult to crack).

Without further discussion of cryptography, but only to explain our white-box watermarking method, we take the remaining nine bits as the result of taking the modulo 512 of the previous 23 bits (or other hash function). Above is illustrated in Figure \ref{figure3}.

The self-check focuses on detecting whether the current parameter has been tampered with. If no error is found in self and mutual check, the probability of misjudgment can be calculated as:
$P{}=\frac{1}{512} \times \frac{2^{23-11}}{2^{23}}=\frac{1}{2^{20}}\approx\frac{1}{1 \times 10^{6}}$.

\begin{table*}
\centering
\setlength{\tabcolsep}{4pt}
\caption{Compares our watermarking method with previous model watermarking methods in object, positioning accuracy, recovery ability, embedding capacity, and embedding method.}\label{Table:2}
\scriptsize
\renewcommand{\arraystretch}{1.6}
\begin{tabular}{cccccc}
\hline\hline
Schemes        & Object                & Localization accuracy & Recoverability & Capacity\cite{reversible} & Embedding method \\ \hline
ACM-\cite{first} & Copyright & -                        & ✗             & Small    & Regularization   \\
ACM-\cite{reversible} & Integrity             & -                        & ✗             & Medium   & Histogram shift  \\
INS-\cite{neunac}  & Integrity             & Block                    & ✗               & Large    & LSB Substitution \\
PRL-\cite{zhao2022dnn}  & Integrity             & Block                    & ✓              & Large    & LSB Substitution \\ 
\textbf{Ours}           & \textbf{Integrity}             & \textbf{Parameters}               & \textbf{✓}             & \textbf{Large}    & \textbf{LSB Substitution} \\ \hline
\end{tabular}
\end{table*}

\begin{table*}
\centering
\setlength{\tabcolsep}{6pt}
\scriptsize
\caption{Compares our method with previous \textbf{fragile watermarking} methods. In embedding stage, the contacting represents modifying parameters and the training means whether the modification is relate to training. The fidelity represents whether the performance of the model remains unchanged before and after modification.
}\label{Table:3}
\renewcommand{\arraystretch}{1.2}
\begin{tabular}{c|cc|cc|cc}
\hline\hline
Schemes          & \multicolumn{2}{c|}{Embedding} & \multicolumn{2}{c|}{Detection} & \multicolumn{2}{c}{Characteristic} \\ \hline
                 & Contact        & Training      & Positioning     & Validator    & Fidelity      & Type                  \\
CVF-\cite{sensitive}   & ✗              & ✗             & ✗               & Trigger      & ✓          & Black-box             \\
KS-\cite{zhurenjie}    & ✓              & ✓             & ✗               & Trigger      & ✗          & Black-box             \\
AAAI-\cite{deepauth} & ✓              & ✓             & ✗               & Trigger      & ✗          & Black-box             \\
ICIP-\cite{yin2022neural}  & ✗              & ✗             & ✗               & Trigger      & ✓          & Score-based Black-box \\
ACM-\cite{reversible}   & ✓              & ✗             & ✗               & Hash         & ✗          & White-box             \\
INS-\cite{neunac}    & ✓              & ✗             & ✓               & Hash         & ✗          & White-box             \\
PRL-\cite{zhao2022dnn}    & ✓              & ✗             & ✓               & Hash         & ✗          & White-box             \\ 
\textbf{Ours}    & \textbf{✓}     & \textbf{✓}    & \textbf{✓}      & \textbf{Hash}         & \textbf{✓} & \textbf{White-box}             \\ \hline
\end{tabular}
\end{table*}

And this approach ensures that when one parameter is damaged, we can choose another parameter for restoration (if there is no self-check, it is impossible to determine which parameter is damaged when checking between parameters). We compared our white-box watermarking with existing watermarking and found that only our method achieves parameter-level positioning and restoration, while achieving optimal performance in fidelity and lossless watermarking, which as shown in Table \ref{Table:2} and Table \ref{Table:3}.

\section{Experiment}
In this section, we selected four classic DNN models, LeNet, AlexNet, ResNet18 and ResNet50 as experimental objects. These DNN models are becoming increasingly larger and deeper, demonstrating the impact of watermark information on different models, and also proving the effectiveness of our adaptive method. We also selected datasets that match the models to make the watermark experiments more practical. The LeNet was conducted on MNIST. AlexNet,ResNet18 and ResNet50 conducted on CIFAR-10. The random seed was set to 1234 for all experiments. In Section 4.1, we will demonstrate the effectiveness of our proposed adaptive method, and in Section 4.2, we will demonstrate the effectiveness of our method against random parameter attacks.

\subsection{Adaptive Ability}
In our experiment, we compared our work with a similar method \cite{zhao2022dnn}, as shown in table \ref{Table:4}. Although they only replaced 12 bits per parameter, there was still some impact on deeper models, while our method replaced 20 bits for each parameter and can still keep the model performance well through adaptation.

\begin{table*}
\centering
\caption{Comparison of our watermarking method with \cite{zhao2022dnn}'s method across four models.}\label{Table:4}
\setlength{\tabcolsep}{8pt}
\scriptsize

\renewcommand{\arraystretch}{1.2}
\begin{tabular}{ccccccc}

\hline\hline
\multirow{2}{*}{Model} & \multirow{2}{*}{Layers} & \multirow{2}{*}{Dataset} & \multirow{2}{*}{Resize} & \multicolumn{3}{c}{Accuracy(\%)}   \\ \cline{5-7} 
                       &                         &                          &                         & Clean & PRL2022 & \textbf{Ours}  \\ \hline
LeNet                  & 7                       & Mnist                    & $1\times28\times28$                   & 97.65 & 97.65 & \textbf{97.82} \\
AlexNet                & 8                       & Cifar10                  & $3\times256\times256$                 & 97.49 & 97.49 & \textbf{98.63} \\
ResNet18               & 18                      & Cifar10                  & $3\times32\times32$                   & 71.89 & 71.88 & \textbf{72.06} \\
Resnet50               & 50                      & Cifar10                  & $3\times32\times32$                   & 73.93 & 73.62 & \textbf{74.01} \\ 
\hline

\end{tabular}
\end{table*}

We divided the model accuracy into three categories: clean model, before and after adaptive model during the process of embedding the watermark. The results are shown in the Table \ref{Table:5}.
\begin{table*}
\centering
\scriptsize
\caption{Accuracy of four models at three stages.}\label{Table:5}
\renewcommand{\arraystretch}{1.2}
\begin{tabular}{ccccc}
\hline\hline
\multirow{2}{*}{Model} & \multicolumn{4}{c}{Accuracy(\%)}                                                   \\ \cline{2-5} 
                       & Clean Model & Before Adaptive & After Adaptive & \textbf{Improvement} \\ \hline
LeNet                  & 97.65       & 97.54           & 97.82          & \textbf{0.28}                 \\
AlexNet                & 97.49       & 97.63           & 98.63          & \textbf{1.00}                 \\
ResNet18               & 71.89       & 64.16           & 72.06          & \textbf{0.79}                 \\
Resnet50               & 73.93       & 58.27           & 74.01          & \textbf{15.74}                \\ \hline
\end{tabular}

\end{table*}
It can be seen that the decrease in accuracy for LeNet and AlexNet after adding the watermark is relatively small, while the decrease in accuracy for ResNet is relatively large. In particular, for ResNet50, the deeper network, the impact of the watermark is even more significant, reaching 15.72\%. We believe this is because the small influence of the watermark on the parameters is amplified layer by layer in models with more layers, leading to a significant performance drop in the end. However, this also confirms the feasibility of our method. We successfully restored the performance of each model to its original level, even with slight improvements. However, it is unrealistic to expect that the adaptive method can significantly improve the model performance beyond the original level.

\vspace{-1.5em}
\begin{figure*}[htp]
	\centering
		\includegraphics[scale=0.23]{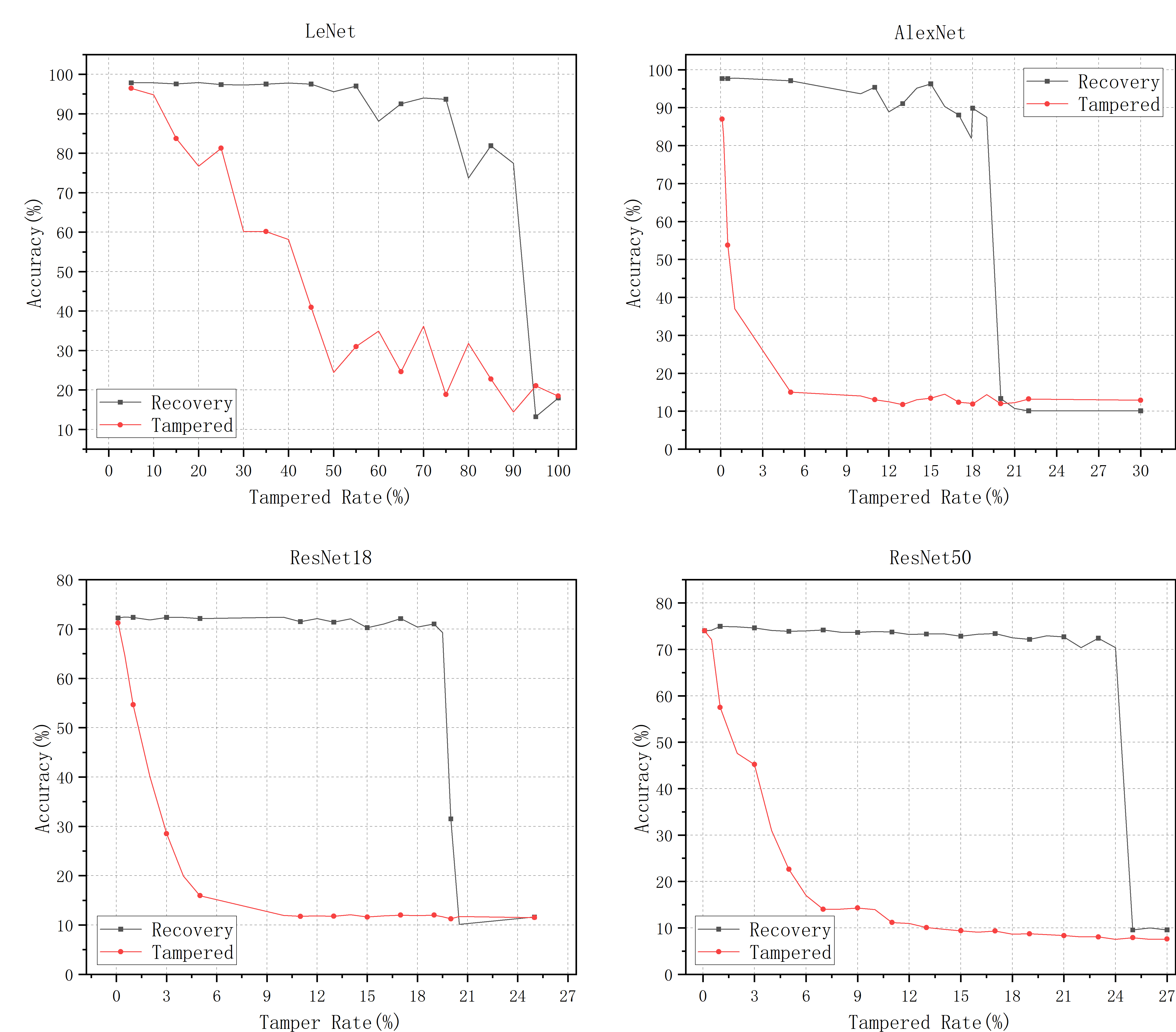}
	\caption{Four models' recovery performance under arbitrary parameter attacks.}
	\label{figure4}
\end{figure*}

\subsection{Performance Recovery}
It has been observed that even small variations in model parameters can have a significant impact on the overall performance of the model \cite{xue2022advparams}, rendering traditional image restoration methods incapable. Therefore, we aim to restore the model's original parameters as much as possible. We assumed attack randomly selects model parameters for random number attacks, and the randomly set numbers will not exceed the size range of the original parameters. In order to compare the performance difference between the attack and the recovery, we select the first layer for testing. It can be seen that our recovery method achieved a uniform decline in performance for the smaller LeNet model, until the attacked parameters reached 90\% and the performance quickly declined. For other larger models, we achieve recovery performance within 20\%. The specific results are shown in the Figure \ref{figure4}.

\section[]{Conclusion}
In this paper, we advocate for the integration of neural network watermarking with the characteristics of neural networks. To achieve this, we propose the application of gradient descent in neural network watermarking, introducing adaptive watermarks. Additionally, we aim to tightly associate each parameter's watermark, carrying the important information of parameters. We propose self-mutual check parameters to enable precise verification and recovery. We combine these two methods and conduct experiments on multiple networks, demonstrating the effectiveness of our approach. The adaptive technique also achieves a significant increase in watermark capacity, allowing for more watermark information to be embedded under lossless conditions in future works.

\bibliographystyle{splncs04}
\bibliography{cas-refs}





\end{document}